# Learning Context: A Unified Framework and Roadmap for Context-Aware AI in Education

Naiming Liu, Brittany Bradford, Johaun Hatchett, Gabriel Diaz,
Lorenzo Luzi, Zichao Wang, Debshila Basu Mallick, Richard Baraniuk

Rice University | OpenStax | SafeInsights | Adobe Research

**Summary**

We introduce a unified Learning Context (LC) framework designed to transition AI-based education from context-blind mimicry to a principled, holistic understanding of the learner. This white paper provides a multidisciplinary roadmap for making teaching and learning systems context-aware by encoding cognitive, affective, and sociocultural factors over the short, medium, and long term. To realize this vision, we outline concrete steps to operationalize LC theory into an interoperable computational data structure. By leveraging the Model Context Protocol (MCP), we will enable a wide range of AI tools to "warm-start" with durable context and achieve continual, long-term personalization. Finally, we detail our particular LC implementation strategy through the OpenStax digital learning platform ecosystem and SafeInsights R&D infrastructure. Using OpenStax's national reach, we are embedding the LC into authentic educational settings to support millions of learners. All research and pedagogical interventions are conducted within SafeInsights' privacy-preserving data enclaves, ensuring a privacy-first implementation that maintains high ethical standards while reducing equity gaps nationwide.

# 1. Introduction

## 1.1 Motivation and Significance

The field of educational technology (edtech) is defined by cycles of innovation, but what we have witnessed with the 2022 release of ChatGPT is not a typical cycle. It is an **arrival technology** [1]. Like the web browser in 1994 and the smartphone in 2007, Generative AI (AI) systems such as Large Language Models (LLMs) have not seen a gradual, top-down adoption; they have triggered a seismic, bottom-up shift in how information is accessed, synthesized, and created. Evidence for this user-driven shift is already clear: a recent analysis by the National Bureau of Economic Research found that “Practical Guidance,” a category explicitly including “Tutoring or Teaching,” is the most common use case for ChatGPT, accounting for 29% of all interactions [2]. Despite this clear, learner-driven demand, AI research and development (R&D) to date have focused primarily on building coding agents rather than teaching and learning agents. AI technology has *arrived*, and its integration into learning is no longer a question of “if” but an urgent question of “how.”

The evidence for this pervasive and organic adoption is overwhelming. K-12 and college students have been the first movers, integrating AI into their learning workflows far faster than any formal curriculum could, with the majority now using AI tools for learning, from homework help and essay drafting to conceptual exploration [3]. This organic use has been corroborated by a wave of recent peer-reviewed studies [4], [5] demonstrating that learners are not just using these tools, but are actively developing new, often unscaffolded, methods for co-learning with AI. Simultaneously, teachers and instructors, initially overwhelmed, are now rapidly adopting AI for their own professional work, primarily to reduce workload in lesson planning, content generation, and differentiation [6], [7].

This is a critical juncture for education. The “arrival” of AI has not presented us with a finished educational solution for learners; rather, it has presented us with a powerful, engaging, and dangerously blind new partner. Today's AI tools are **context-blind**,[1] responding to user requests anew as they build short-term memory windows. They are powerful stochastic parrots, capable of generating fluent text but fundamentally incapable of understanding the learner. They have no model of the student's **prior knowledge** or **misconceptions** (a constructivist view), their **affective state** (e.g., “state” anxiety vs. “trait” anxiety), their **social environment and identity** (a socio-constructivist view), nor the **distributed cognitive system** of which they are only one part. And every time a learner switches from one AI tool to another, what little context was acquired is lost. AI cannot differentiate between a curious learner exploring a concept and a struggling learner reinforcing a deep misconception, and learners themselves often cannot articulate the support they need for success.

[1] Here “context-blind” refers to the absence of durable, educationally meaningful context. While modern LLMs maintain a finite context window for short-term text recall, this transient buffer does not capture or retain a learner’s cognitive, affective, or sociocultural state across sessions or tools.

Realizing the true promise of AI for education will require a community-wide, multidisciplinary **research and development (R&D) program** that makes AI tools **context-aware** for all learners. Such systems will promote deep, robust learning rather than optimizing for shallow, plausible-sounding mimicry, ensuring that this powerful "arrival technology" closes equity gaps instead of widening them. In this document, we map out a roadmap for such a program and overview the initial steps toward its realization being taken at OpenStax [12] and SafeInsights [13]; we invite the broader R&D community to join us in this collaborative effort to build a more personalized, equitable, and evidence-based learning ecosystem.

### 1.2 A Roadmap for Learning Context R&D

The high-level goal of our proposed R&D program is to **make AI-based teaching and learning systems context-aware** by developing a **Learning Context (LC)** that encodes the key factors relevant to a learner's learning state over the short, medium, and long term. Access to the appropriate elements of the LC will enable virtually any AI tool to "warm start" rather than "cold start" [111] and more ably and efficiently guide a learner towards deeper learning. The R&D program leverages the burgeoning progress on context in the AI agent space [8], [9] that has not yet arrived in education and is naturally organized into four interconnected themes that integrate teaching and learning and technology research.

**Theme 1: Learning Context Theoretical Framework.** Although AI systems excel at content delivery and procedural support, they generally neglect the cognitive, emotional, and sociocultural dimensions that shape authentic learning [10], [31]. Drawing on decades of learning science research showing that learning is deeply contextual and dynamic [11], we propose an initial LC theoretical framework that represents and leverages these multifaceted dimensions of learning.

**Theme 2: Learning Context Technology.** The LC theoretical framework can be translated into a computational agenda, resulting in efficient, machine-readable, human-interpretable data structures that can be leveraged by AI systems to provide more personalized teaching and learning experiences. Research here should focus on four key tasks: LC representation, LC compression/prioritization, LC tracking, and LC utilization.

**Theme 3: Application, Testing, and Validation.** To bring together different stakeholders to co-design the LC and implement it in AI instructional tools, we propose a design-based implementation research (DBIR) framework blended with user-centered design (UCD). Testing and validation of this approach at OpenStax and SafeInsights involves integrating LC-based AI tools into the new OpenStax Assignable teaching and learning platform [12] and studying learner engagement and learning outcomes in a privacy-preserving fashion via integration with SafeInsights [13], the US national infrastructure for learning and education research.

**Theme 4: Privacy and Ethics.** The LC holds considerable promise for improving AI-supported teaching and learning, but it also presents significant privacy risks that must be addressed intentionally. To ensure success across all research themes, it is critical to engage a diverse group of stakeholders—particularly students and instructors—as active partners in a User-Centered Design (UCD) process. This collaborative effort will address fundamental questions regarding the ownership and hosting of the LC, while carefully navigating the essential tradeoff between data privacy and instructional informativeness.

Making progress on the above agenda will require a **multidisciplinary research community** that draws expertise from AI systems, the learning sciences, discipline-based education research, human-computer interaction, privacy, and ethics, among others.

### 1.3 Learning Context Vignettes

We demonstrate through four simple experiments that, for learning, context matters.

**Vignette 1. Student context personalizes AI instruction.** AI tutoring can adapt its instructional approach when provided with a different learner's context. As we see in Figure 1, given the same mathematics word problem, the AI tutor produces two distinct explanations: one for **Alex**, a learner with strong language skills but weaker in mathematics, where the solution is framed through narrative and concept-building; and another for **Blake**, who is strong in mathematical reasoning but has weaker language comprehension, where the explanation is concise, equation-driven, and math-forward (see the learner personas below). This contrast highlights that effective learning support is not one-size-fits-all and **instruction must be aligned with the learner's prior knowledge, cognitive strengths, and background**. By tailoring responses to these contextual factors, we see how context-aware AI can deliver more accessible, engaging, and equitable learning experiences.

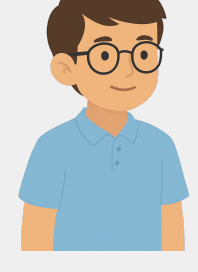

**Alex** performs strongly in English, especially in creative writing and understanding stories, but finds math more challenging. He often freezes on multi-step problems and has trouble turning them into equations. Alex is highly verbal, enjoys storytelling, and responds well to supportive, low-pressure guidance. Positive reinforcement helps reduce his anxiety and keep him engaged.

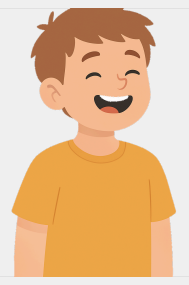

**Blake** is a quick thinker who loves puzzles, patterns, and figuring things out before anyone else. Math feels like play to him, and he enjoys turning tricky problems into challenges he can beat. Reading, on the other hand, slows him down, especially when wordy stories hide the numbers he needs. When learning feels active and visual, Blake's curiosity and confidence shine.

**Vignette 2. Context induces a learner-centered shift in LLMs' instructional priorities.** To test if an LLM can adapt its instructional strategy to a learner's needs, we conducted an experiment where an LLM was tasked with selecting the top three teaching strategies for **Maya**, an anxious learner studying calculus.

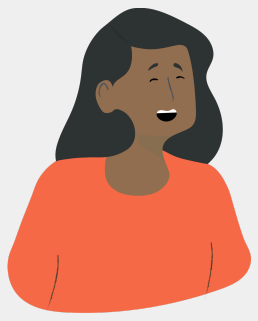

**Maya** is a first-year college student struggling to stay motivated and confident in her coursework. She often feels disengaged from the material and doubts that her efforts will lead to mastery. When studying, she tends to feel overwhelmed, often giving up before completing tasks, especially when concepts become confusing. Exams are a particular source of anxiety, as she frequently compares herself negatively to others.

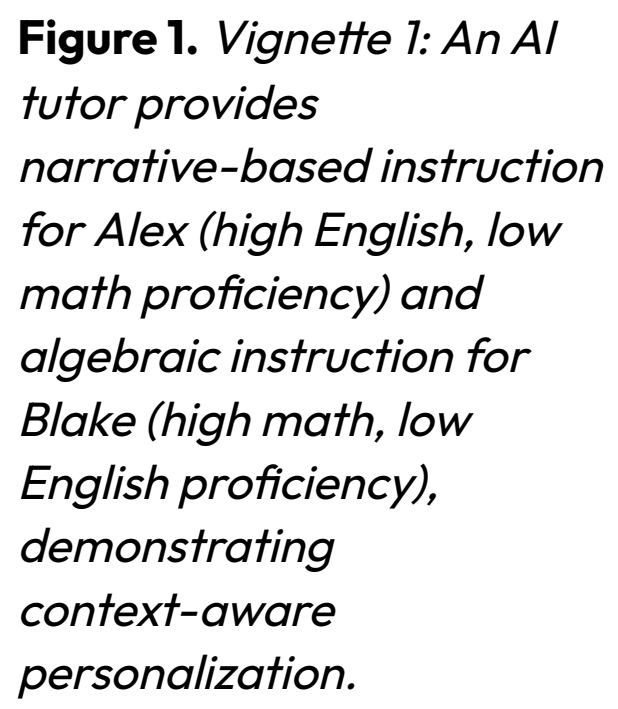

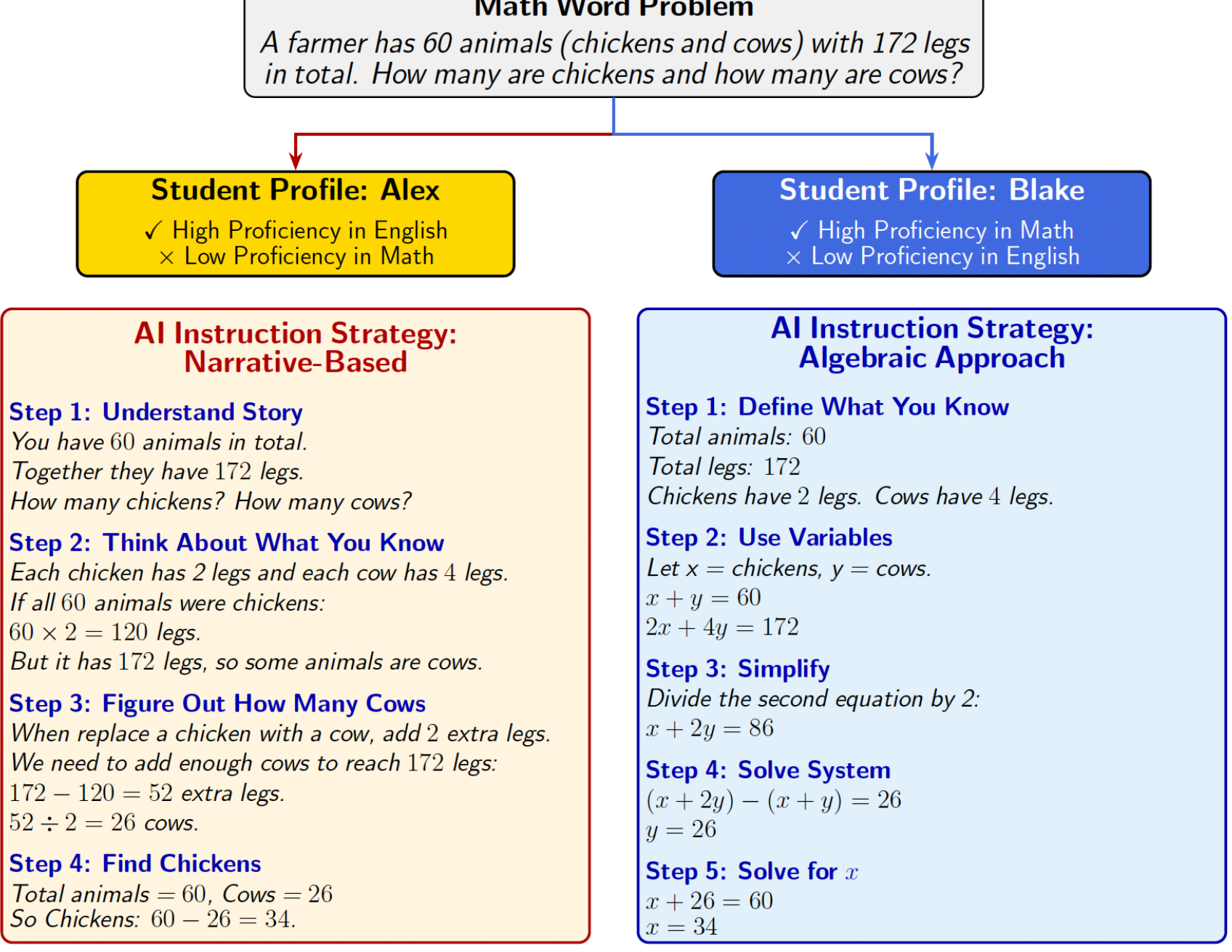

**Figure 1.** *Vignette 1: An AI tutor provides narrative-based instruction for Alex (high English, low math proficiency) and algebraic instruction for Blake (high math, low English proficiency), demonstrating context-aware personalization.*

The experiment was conducted with two conditions: a control group with no information about Maya and a treatment group where the LLM was given Maya's profile. The results, summarized in Figure 2, show a significant and pedagogically relevant shift in the LLM's instructional choices. In the absence of context about Maya's learning characteristics, the model consistently selects content-focused strategies suitable for teaching calculus, such as "guided practice" and "worked examples".

In contrast, when provided with Maya's profile, the model's pedagogical strategy shifted to address her specific needs. It selected "foster growth mindset" in every trial and "goal setting and monitoring" in nearly every trial—direct and appropriate responses that target the stated self-efficacy and anxiety barriers. This result demonstrates that **even minimal learner context can cause an LLM to shift its instructional priorities from generic and content-first to specific and learner-centered**. Future LC work can build on this foundation to develop architectures for personalizing education at scale, creating AI tutors that can reason about *how* to teach, not just *what* to teach.

**<u>Vignette 3. Current AI suffers from a relevance-impact misalignment</u>.** To determine whether current AI models can natively distinguish between critical learner signals and noise, we measured the distributional shift (via the Total Variation Distance, with smaller distance indicating less shift) induced by injecting low-relevance learner characteristics (e.g., "The learner has a collection of interesting rocks and minerals") into the learner context. We hypothesized that highly relevant characteristics (e.g., a learner's degree of self-efficacy) would drive significant shifts in the model's instructional strategy, while irrelevant characteristics would be ignored.

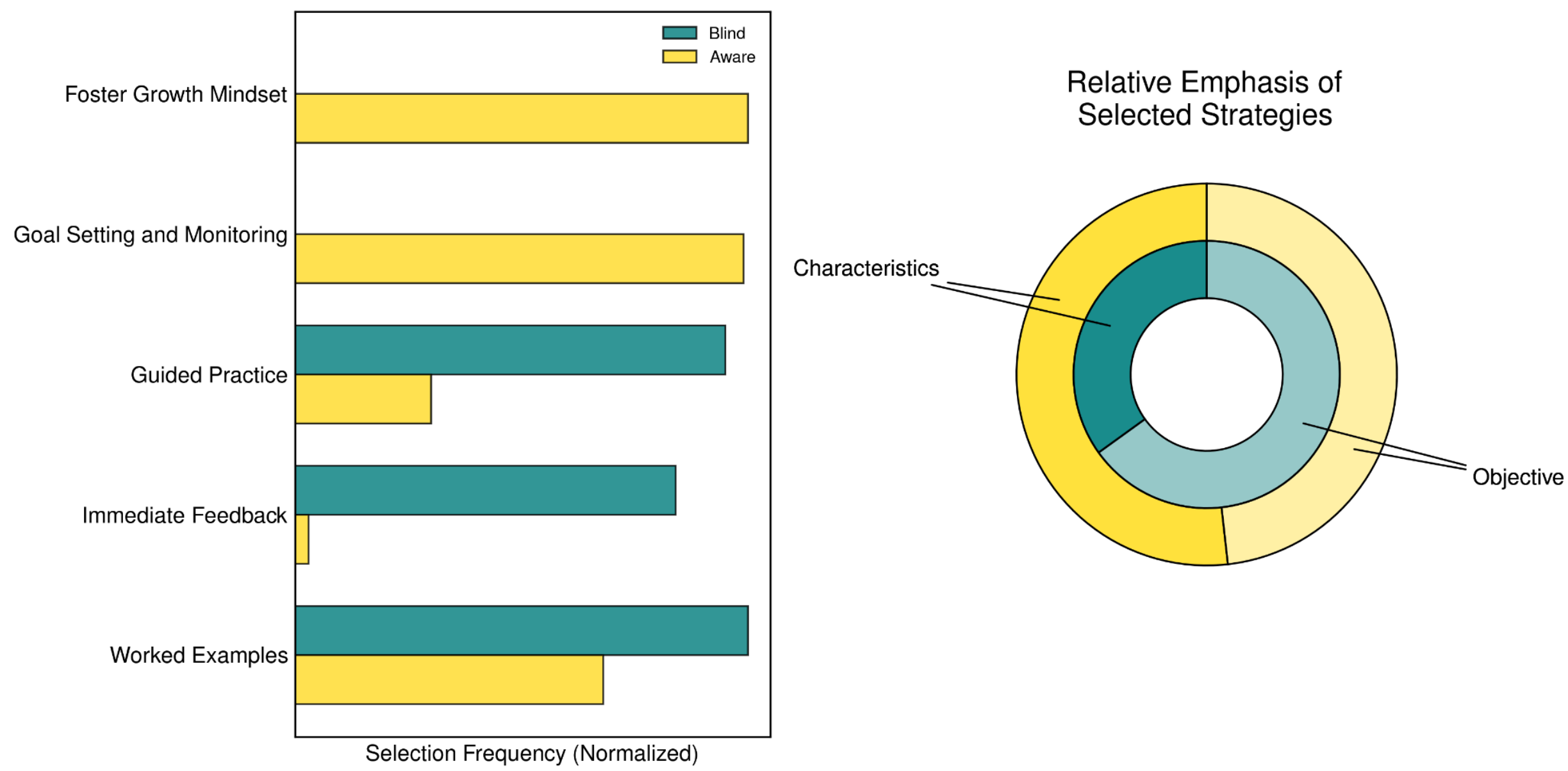

***Figure 2.*** *Vignette 2: Impact of learner context on instructional strategy selection. (Left) Frequencies of pedagogical strategies generated by an LLM in context-blind vs. context-aware instructional design scenarios for an anxious learner. Displayed categories represent the union of the top 3 strategies from both conditions. (Right) The same strategy counts aggregated by relative pedagogical emphasis, illustrating a qualitative shift from general content-focused instruction to characteristic-specific support when the learner context is provided. Pedagogical strategies are derived from the Digital Promise Learner Variability Navigator [14].*

To quantify the individual contribution of each characteristic in the LC, we employed a leave-one-out (LOO) analysis: for each characteristic, we generated a variant context with that characteristic omitted and then prompted an LLM to execute the strategy selection task in Vignette 2. By computing the Total Variation Distance (TVD) between the full-context strategy distribution and each LOO-context distribution, we measured the marginal impact of each characteristic. This approach assesses the degree to which the model's sensitivity to each characteristic aligns with that characteristic's pedagogical relevance.

The results in Figure 3 reveal a critical alignment gap. While the model correctly prioritizes impactful traits like the learner's perceived value of the task (TVD: 0.287) and self-efficacy (TVD: 0.207), it fails to attend to other vital factors such as the learner's ability to regulate their effort (TVD: 0.080), treating them as effectively invisible. Conversely, the model exhibits "hallucinated relevance," assigning a meaningless distractor (i.e., favorite hobby) a higher impact weight (TVD: 0.273) than nearly all pedagogical traits. This significant variability underscores that **we cannot rely on current AI systems to natively prioritize factors in the LC and motivates the context prioritization framework** we outline below in Section 3.3 in order to compress and select relevant context features before they influence AI responses.

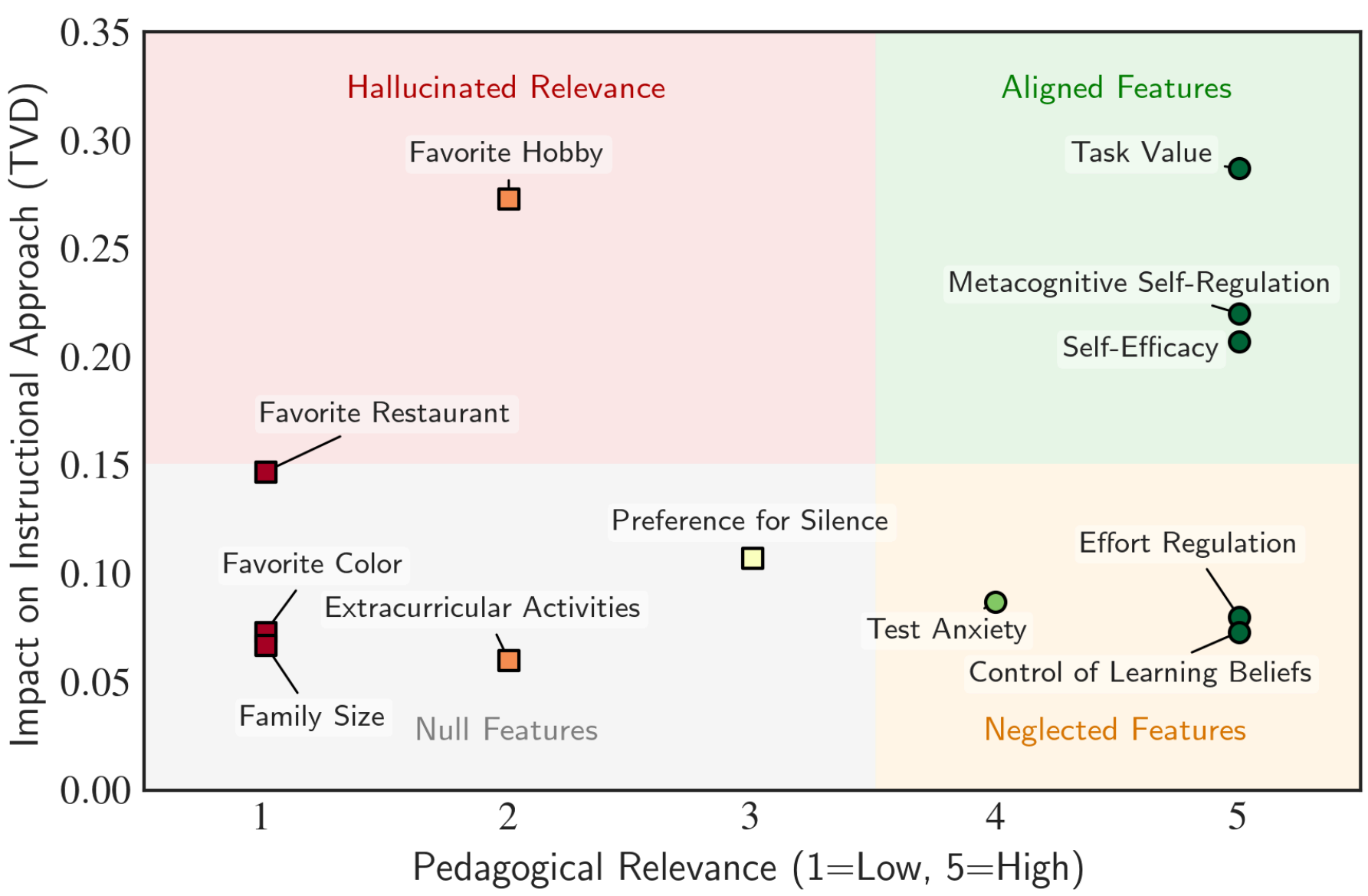

**Figure 3.** *Vignette 3: The relevance-impact misalignment in AI. The lack of a linear relationship indicates a failure in implicit context prioritization. Ground-truth learner characteristics (circles) are sourced from the MSLQ [15], and distractors (squares) are synthetically generated.*

**<u>Vignette 4. Closed-loop validation of learning context</u>.** To evaluate the feasibility and internal validity of the LC framework, we designed a closed-loop experimental pipeline (Figure 4) that jointly tests whether LLMs can (i) simulate high-fidelity, context-driven interactions conditioned on a specific LC and (ii) recover that same LC from student–tutor interactions alone. The core question is whether the LC is consistently preserved across this generation-recovery loop.

The pipeline employs the following validation process:

- **Context-conditioned interaction simulation:** We generated multi-turn student–tutor dialogues using an LLM conditioned on an explicit LC comprising two components: cognitive misconceptions (underlying belief, erroneous example, and triggering feature) and learner profile attributes (anxiety, conscientiousness, and language proficiency). The AI tutor was required to manifest these components naturally within the conversation.
- **Learning context recovery from interactions:** We reconstructed the learner's LC using a second LLM (blinded to the original LLM and its input) by inferring student misconceptions and profile attributes directly from the dialogue, supported by quoted evidence.

- **Self-validation loop:** Because the input and recovered LCs share a common schema, we can directly compare them to assess **LC consistency**. This loop tests whether LC specifications are faithfully preserved during generation and reliably observable from dialogue.

In a pilot study of 35 simulated interactions with the GPT-5.1 model, we observed a significant "observability gap" between conceptual and behavioral data:

- **Reliable conceptual recovery:** The underlying misconception was correctly identified in 91.4% of cases. This indicates that belief-level misunderstandings can be consistently preserved and decodable via dialogue traces.
- **Differential profile observability:** Recovery of learner attributes varied significantly. While anxiety was recovered with 100% accuracy, conscientiousness (68.6%) and language proficiency (60%) were substantially more difficult to infer.

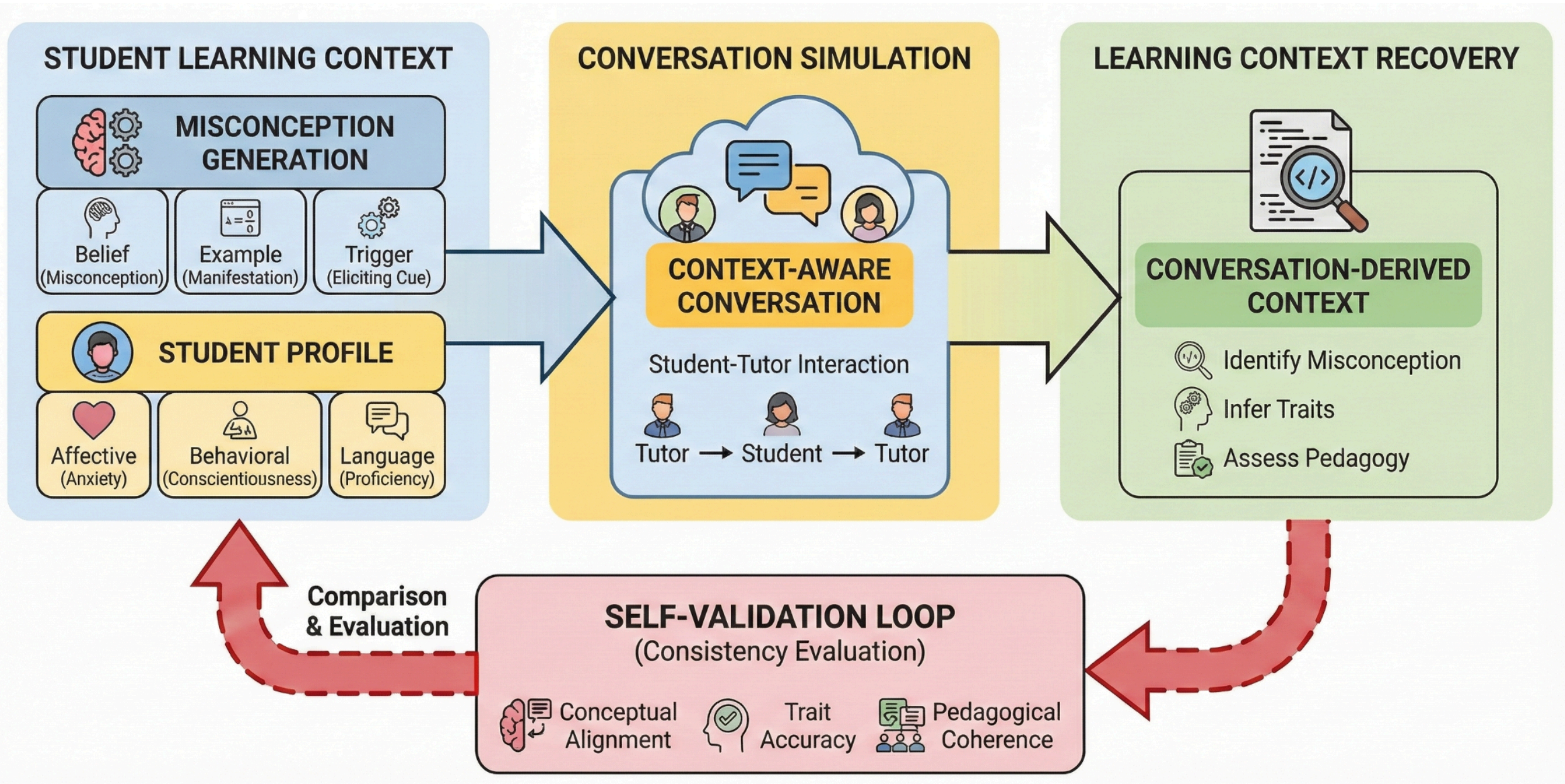

**Figure 4.** *Vignette 4: LC self-validation pipeline that measures the fidelity of LC preservation.*

These results demonstrate that the **LC is a measurable, durable representation capable of preserving vital learner signals across different AI platforms**. Crucially, the finding that behavioral and linguistic traits (such as conscientiousness and language proficiency) can take significantly longer to recover reliably than affective states provides the primary motivation for the LC data layer. **By capturing these "slow-to-surface" traits, the LC will enable AI tools to "warm-start" with a rich, pre-existing understanding of the student**. This prevents the "cold-start" problem where a learner must spend valuable time re-proving their unique needs to every new AI system they encounter, enabling immediate, personalized support from the very first interaction.

## 2. Theme 1: Learning Context Theoretical Framework

Two core theoretically grounded R&D questions drive this theme: (i) How do established learning science theories conceptualize LC, and what commonalities do they share? (ii) How can we represent a learner's holistic, dynamic context in a way that is grounded in learning science, computationally amenable, machine-readable, and human-interpretable?

### 2.1 Context as Foundational to Learning Theory

Contemporary AI-supported learning systems incorporate **context that is limited to users' interactions with the AI system**. Although these systems excel at content delivery and procedural assistance, they largely ignore the cognitive, affective, and sociocultural factors that shape learning in real settings [16]. This limitation stands in contrast to decades of learning science research demonstrating that learning is not a universal, decontextualized process, but deeply contingent on learners' evolving mental states, prior experiences, goals, and environments [17]. Without access to this contextual information, AI tutors and educational agents cannot meaningfully personalize support or learning. Further, learners frequently learn across multiple tools and settings [18]. Treating their learning as static, homogeneous, and tool-isolated contributes to inequity and limits the potential of AI to support users' full learning journeys.

### 2.2 Cross-Theory Representations of Context as an Active Constituent of Learning

Although operational definitions vary, **context** is consistently positioned as a core element of learning across multiple theoretical frameworks, not a peripheral background variable. We draw from these domains to illustrate the centrality of context and to ground our expansion of context into an AI setting.

**Learning contexts are sociocultural.** In *sociocultural theory* [19], LC comprises the tools, symbols, interactional patterns, and cultural norms through which individuals participate in shared activity. Higher mental functions emerge first between people and only later become internalized. This framework also informs the *zone of proximal development*, the distance between what a learner can accomplish independently and what they can achieve with scaffolded support from a skilled instructor. Participation in cultural practices and their modalities (e.g., language, diagrams, mathematical notation, digital tools) supports meaning-making [19]. Similarly, in *situated learning theory*, knowledge is inseparable from the activity, culture, and social setting in which it develops [20], [21]. Learning occurs through participation in communities of practice, where norms, tools, language, and shared goals shape meaning-making. Although *constructivism* emphasizes individual experiences, it also recognizes that learners interpret experiences within social and cultural settings [22].

**Learning contexts include tools, artifacts, and representations.** In constructivist theory [22], learning is strengthened when individuals construct external artifacts that make thinking visible and revisable. *Distributed cognition theory* [23], [24] similarly emphasizes that learning depends on the organization and use of social and material resources. In *ecological systems theory* [25], learners experience recurring interactions with people, objects, and symbolic tools within both immediate settings (i.e., the microsystem) and across broader environments (e.g., home, school, online) via proximal processes.

**Learning contexts are interconnected.** In *activity theory* [26], cognition, tools, goals, and social relations form a single system of activity that defines the context of learning. *Distributed cognition theory* extends the cognitive system to include people, artifacts, representations, and the physical environment. Learners co-construct knowledge and problem-solving capacities through interaction with representational tools, social partners, and physical environments [19], [23], [27], [28].

**Learning contexts are dynamic and changing.** In *activity theory*, learners move across institutional contexts, tools, and communities to assemble resources and knowledge in boundary-spanning behaviors [26]. In *constructivist theory* [22], learning is an active process in which individuals build understanding by interacting with their environment, and context is the experiential space in which meaning-making occurs (e.g., problems encountered, tools engaged with, and feedback received). Meaningful learning requires authentic tasks, opportunities for experimentation, and structured reflection, because context shapes what learners attend to and how they interpret it [20], [29]. In *distributed cognition*, context drives how information flows, how representations are used, and how problem-solving unfolds [21], [30].

These perspectives establish the principles upon which the R&D community can build a framework for LC in AI to encompass sociocultural elements, including learner characteristics and experiences; tools and related resources; as well as reflect the reality that these elements are interconnected and dynamic.

### 2.3 Learning Context Framework for AI Applications

Modern learning occurs across platforms, social spaces, and modalities, yet **current AI systems capture only the context visible through direct user interaction**. Thus, they operationalize only a narrow subset of the factors learning science identifies as central to learning. Despite broad recognition of context's importance [32] , the field lacks a computationally tractable theory for representing LC at scale.

To bring learning science into AI-supported education, formalize and operationalize contextual dynamics, and encode a broader ecology of learning, we propose a unified computational framework for representing and leveraging LC. Rather than positioning AI as a standalone tutor, we reconceptualize it as part of a **human-AI partnership** that adapts to the holistic learner and captures cognitive dynamics, affective conditions, social settings, and tool-mediated learning.

This paradigm shift informs the development of AI models that learn with learners over time, travel across learning tools, and support adaptive autonomy rather than rigid control.

To orient our research agenda, we propose the **Learning Context Framework for AI Applications** ("The LC Framework" for short; see Figure 5) that structures context across five core dimensions:

| Dimension | Description | Construct Examples |
|---|---|---|
| **"Who"** | Learner factors | Intra-individual (e.g., prior knowledge, misconceptions, metacognition) |
| **"With whom"** | The peers/instructors that learners interact with | Inter-individual (e.g., peer collaboration, instructor feedback) |
| **"What"** | The content | Course content (e.g., problem difficulty, domain (STEM vs. non-STEM)) |
| **"When"** | Temporal element, stability over time, state vs. trait constructs | Stable traits (working memory; reading skill) vs. volatile states (cognitive load, affect, motivation) |
| **"Where"** | Digital environment, based on Bronfenbrenner's [25] ecological levels | Micro-system (e.g., immediate task, task structure) to meso-system (e.g., between AI environments) |

**<u>Contextual dimensions</u>.** Each of the constructs mentioned in these five dimensions represents a subset of the relevant constructs that AI learning tools can capture (Figure 5, left).

***Learner factors ("Who").*** *Background.* Background characteristics such as age and English learner status predict learners' engagement and performance [33]. *Prior knowledge.* Students with a stronger knowledge foundation in a domain often have better learning outcomes when learning new topics in that domain, although several mediators (e.g., learner characteristics, learning environment) also impact the relationship [34]. *Psychosocial traits.* Students high in certain personality traits in the Five Factor model, especially conscientiousness, tend to have higher academic performance [35]. Conversely, students high in anxiety tend to perform worse than otherwise similar students [36]. *Cognitive factors.* Working memory capacity, or executive attentional control, predicts academic performance [37].

***Peer and instructor factors ("With whom").*** *Peer collaboration.* High-performing students are more likely to engage in supportive peer behaviors and foster knowledge sharing in online learning contexts, such as discussion boards [38]. *Instructor feedback.* Although corrective feedback is critical for fostering learning, motivational feedback is underutilized despite its positive impact on learners' affective states and engagement in online settings [39]. *Instructor engagement and attitudes.* High-quality instructor engagement with learners in asynchronous online classes predicts greater learner satisfaction and achievement [40], and instructor openness to change predicts technology integration [41].

***Content ("What").*** *Task difficulty and complexity.* Task difficulty reflects a learner's subjective perception of challenge, whereas task complexity denotes the inherent cognitive demands of the task [42]. Successfully completing difficult tasks requires greater learner motivation and self-regulation, but also predicts better learning outcomes. *Domain.* Different subject areas require different levels of proficiency, representational competencies, and foundational knowledge [43], [44].

***Digital environment ("Where").*** *Micro-level.* Online tasks range from completing multiple-choice quizzes in an LMS that are automatically graded to instructor-graded free-response exams to online interactions with peers [45]. *Meso-level.* To be successful in their learning journeys, learners must also demonstrate *digital fluency* as they navigate between multiple online learning environments [46].

***Time ("When").*** *State vs. trait constructs.* Traits are stable, long-lasting characteristics that shape behavior across settings and over time (e.g., conscientiousness). In contrast, states are temporary, context-dependent psychological conditions (e.g., math anxiety) that can shift rapidly [47]. Moreover, individuals themselves change, blurring the line between states and traits as a result of longer-term shifts in disposition [48]. *Intraindividual changes.* Within-course intraindividual changes, such as how a learner's engagement rises or dips in response to different instructional approaches, or how patterns of peer support and social interaction develop across a course, predict longer-term learning outcomes [49].

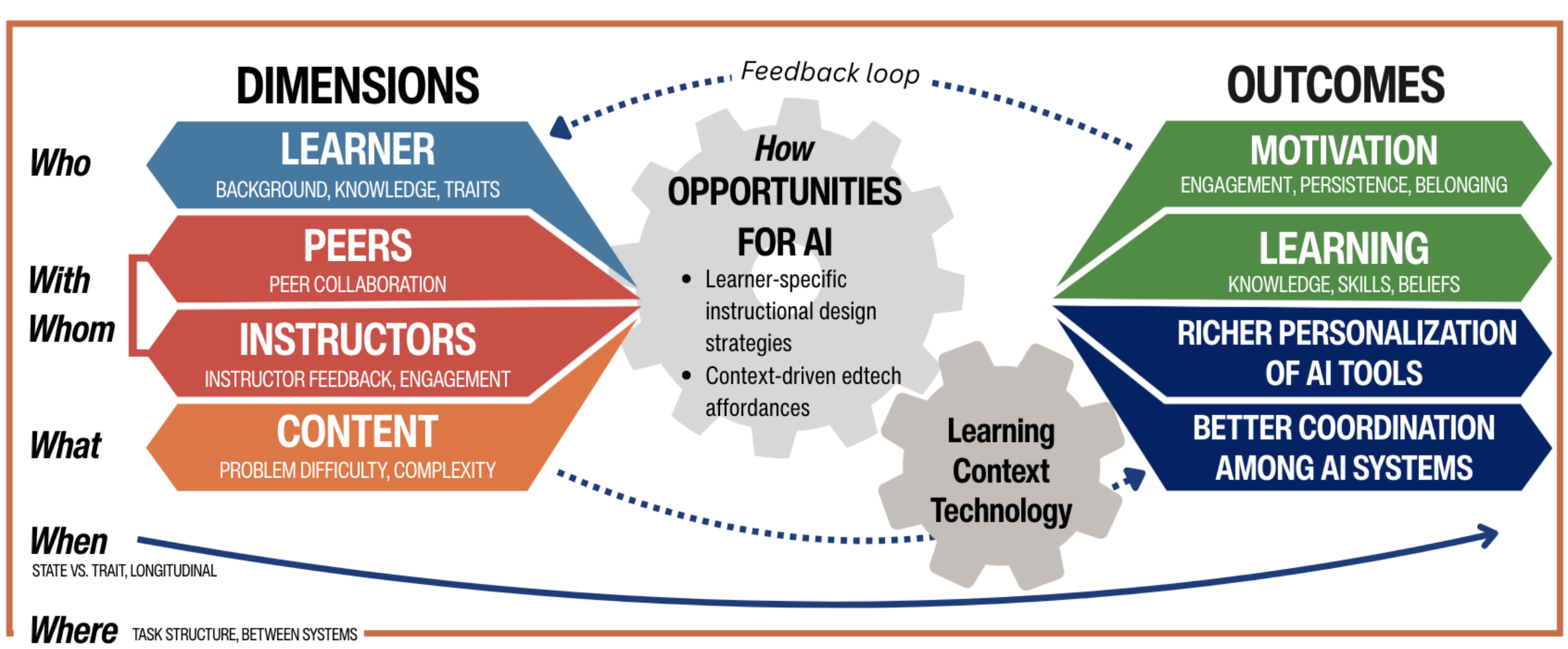

**Figure 5.** *Learning Context (LC) framework for AI for education applications.*

**Opportunities for AI (the "How").** Jointly, these five contextual dimensions drive our **theoretically grounded synthesis of opportunities for integration with AI capabilities** (Figure 5, center*).* These opportunities are grounded in prior research on the impact of contextualizing instructional strategies [50] in online learning contexts, as well as the affordances of educational technology in offering learning-science grounded strategies.

***Contextualized instructional design strategies.*** *Zone of proximal development.* In adaptive learning platforms, learners perform better when they are given assignments that match their "ready to learn" profiles [51]. *Task difficulty*. Matching task difficulty to a learner's current knowledge base predicts better learning outcomes [51], and just-in-time support can bolster learners' persistence when solving challenging problems [52]. *Adaptive sequencing.* Rather than presenting a fixed sequence of activities or lessons to all learners, adaptive sequencing systems use learner modeling to personalize content based on individual learners' progress and performance [53]. *Personalizing the domain.* Personalizing content around a learner's interests supports greater engagement and motivation [52], although attempting to increase interest (e.g., via visuals) can inadvertently produce extraneous cognitive load and decrease performance if the material is not also highly relevant (i.e., the coherence principle [54]). *Managing cognitive load.* One goal of effective instructional design is to manage learners' cognitive load by ensuring the essential processing ability required to master a concept does not exceed the learner's cognitive ability and by minimizing extraneous elements that add to the learner's cognitive load [54].

***Edtech affordances.*** *Feedback.* Receiving personalized, automated feedback improves performance [55]. *Prompts.* Prompts designed to induce self-regulated learning strategies and/or problem-solving strategies in online learning settings lead to more effective task-solving behaviors and increased performance [56]. *Practice quizzes.* Taking online practice quizzes improves learner exam performance for both repeated and novel items [57].

This framework enables the development of AI models that can reason not just about what a learner knows, but with whom, when, where, and under what conditions learning unfolds, enabling personalization that maximizes learners' engagement and performance. As we discuss next in Research Theme 2, the above theoretical framework can be readily translated into new **learning context technology**: a machine-readable representation of LC that enables new context-aware AI systems that can perceive, reason about, and adapt to the evolving activity systems in which learners participate.

## 3. Theme 2: Learning Context Technology

This theme operationalizes the LC Framework introduced in Section 2, which defined the cognitive, affective, and sociocultural dimensions of context across five organizing questions (What, Who, With whom, When, Where). Here, we specify how these constructs will be represented, prioritized, updated, and applied within AI-powered teaching and learning systems. Our objective is to **translate the LC theoretical foundation into a computational framework**: efficient, machine-readable, and human-interpretable data structures that enable AI systems to incorporate context rather than treating all learners identically (Figure 5, center right).

This theme serves as a bridge between learning science theory (Theme 1) and educational application (Theme 3). Each subsection translates core theoretical constructs into a concrete research question. The overall deliverable is a flexible LC architecture that can answer the following research questions: (i) What essential contextual information about learners must be captured? (ii) How can LC be compressed and updated efficiently? (iii) How should LC guide pedagogical actions? and (4) How can modern AI systems interoperate securely through open standards such as the Model Context Protocol (MCP)?

### 3.1 Learning Context Data Structure

With the LC Framework (Theme 1) as our theoretical base, we can translate contextual dimensions to data structure. Each LC dimension is implemented as a structured computational component that, together, form a unified representation of a learner's evolving state. Figure 6 illustrates four approaches to represent LC each with distinct trade-offs.

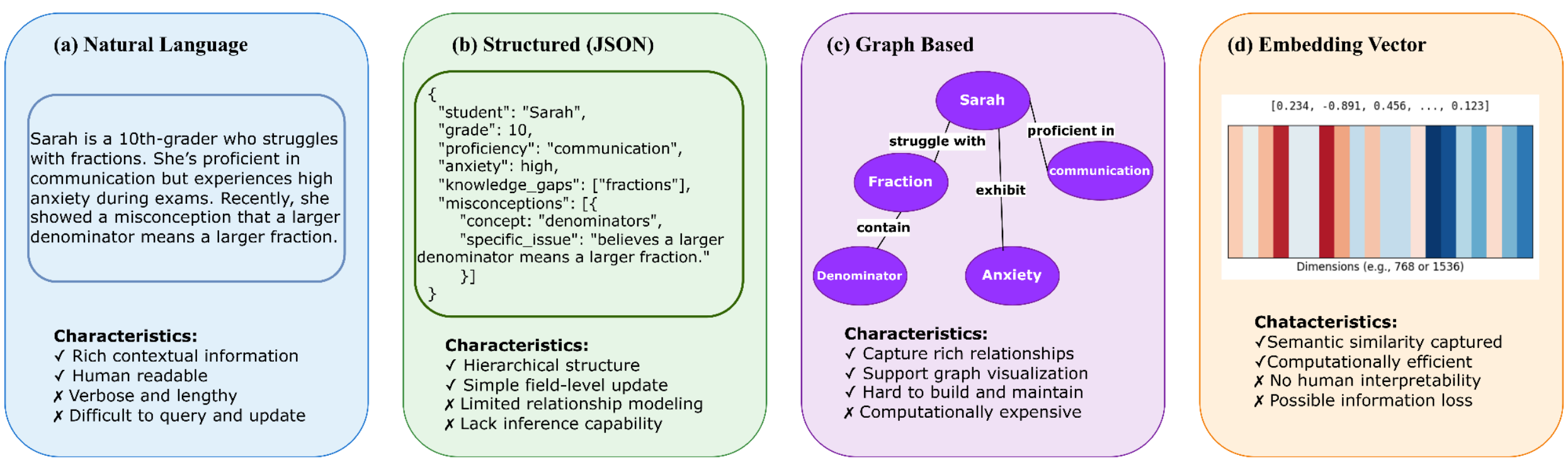

**Figure 6.** *Different representation methods (with examples) for LC.*

**<u>Who</u>.** The **learner model** represents the learner's evolving cognitive, affective, and behavioral profile, including relevant **learner outcomes** (Figure 5, upper right). The LC learner vector combines features derived from edtech affordances (e.g., self-report survey instruments; Figure 5, center) to assess learners' motivation (e.g., affective signals, detected confusion, curiosity, engagement) and learning outcomes (e.g., accuracy, change in knowledge states). Our prior work on learner modeling [106]-[108] has also demonstrated effective strategies for capturing learner cognition. These components are embedded in a latent space informed by psychometric and learning-science theory, ensuring interpretability while allowing adaptive inference.

**<u>With whom</u>.** The **social and collaborative context** is modeled as a dynamic graph where edges represent personal relationships, communication frequency, and collaboration quality. This structure enables AI systems to model social learning dynamics and align with socio-constructivist theory [19].

**What.** The **learning content** and **task context** encode the semantic and hierarchical structure of materials the learner interacts with. We use graph-based representations linking content items to skill and concept taxonomies, employing embeddings aligned to edtech data standards such as IEEE LOM and xAPI [58]. Each node is annotated with metadata describing difficulty, modality, and prerequisite relationships, mirroring the theoretical dimension of the LC.

**When.** The **temporal evolution** captures session timing, spacing of practice, and longitudinal exposure to concepts. The LC temporal index incorporates recency weighting and spacing intervals that approximate human memory decay and retrieval patterns.

**Where.** The **environmental** and **platform context** records the digital environment of learning, including the learning task, modality of content, and learning platform in use. This structured metadata enables models to make situationally informed pedagogical decisions.

**Architectural principles.** These five components form a multi-layered LC graph, where nodes correspond to entities (learner, content, peers, activities) and edges encode relationships and interactions. The graph can be structured as a JSON-LD schema [59] or a knowledge graph compatible with graph databases such as Neo4j [60], enabling both relational and embedding-based reasoning. Prior work in automatic knowledge graph construction (e.g., [61], [62]) and reasoning can inform efficient techniques for representing and using LC data. Our recommended design principles include:

- **Modularity:** Each LC dimension forms an independently addressable module, enabling selective access and updating.
- **Extensibility:** New dimensions (e.g., motivation, goal orientation) can be added without schema overhaul.
- **Interpretable embeddings:** Human-readable "context snapshots" summarize the LC state for instructors and learners.
- **Cross-system consistency:** Standardized metadata ensures interoperability across AI tools and learning platforms.

This representation directly operationalizes the theoretical insight that **learning is situated**: it depends on the interplay among learner, content, and environment. When integrated with AI tools, the LC enables warm-start reasoning: rather than beginning with no prior knowledge of a learner, each new AI system can immediately adapt its responses using LC information accumulated from previous interactions.

**Belief-based models (BBMs).** Certain key components of the LC framework are not yet (fully) developed. For example, we are working towards a BBM component that captures what learners actually believe, such as the mental models, assumptions, preferences, and misconceptions that shape how they think, learn, and engage with content. BBMs address a crucial gap: what learners believe fundamentally shapes their learning trajectories in ways that modeling factual knowledge alone cannot capture, and these beliefs must be represented as part of the broader LC. An example of this integrated structure is shown in Figure 7.

### 3.2 Learning Context Measurement Infrastructure

The extent to which an AI for education system can personalize instruction is fundamentally limited by what it can observe about the learner's context. **Building infrastructure to capture the rich LC data is a cornerstone of modern personalized learning**. However, doing so is challenging: many contextual signals are hard to collect reliably (e.g., facial expression or physiological sensors can be clumsy to deploy), and extensive data collection raises serious privacy concerns. Even so, researchers are making strides in both what kinds  of context data can be captured and how it can be scalably collected, enabling us to move beyond the narrow signals traditionally logged by learning platforms.

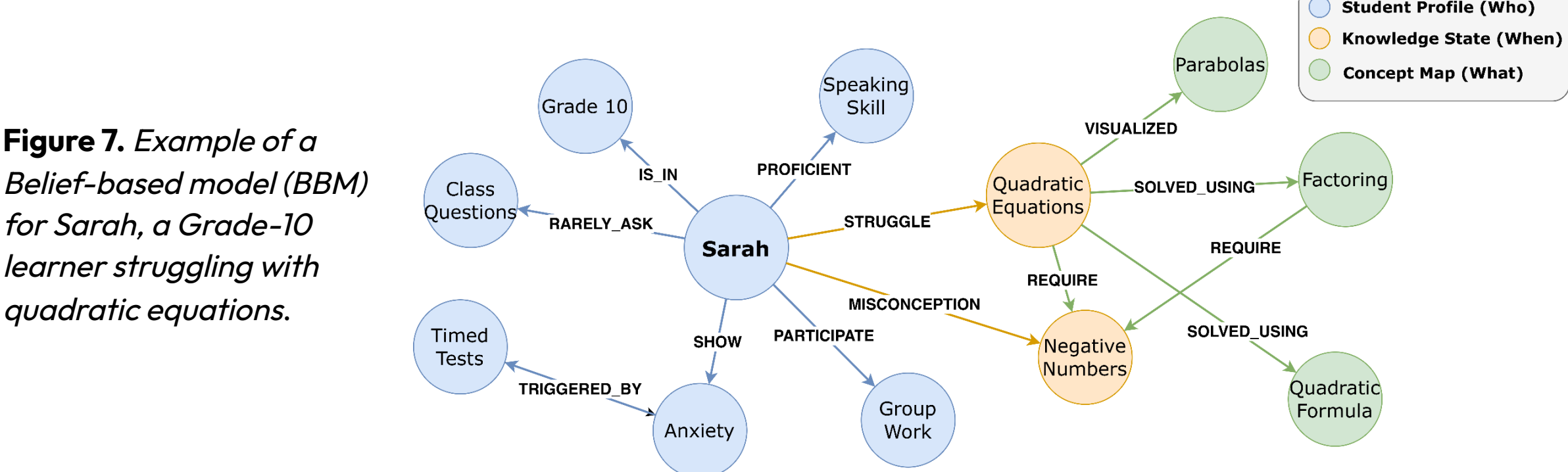

**Figure 7.** *Example of a Belief-based model (BBM) for Sarah, a Grade-10 learner struggling with quadratic equations.*

**Knowledge tracing** (KT) offers a classic example of context data in education. KT models track a student's mastery of skills over time based on their responses to questions. For decades, these models have relied on the simplest context data: whether each answer was *correct or incorrect*. The Bayesian knowledge tracing framework [63], for instance, pioneered modeling of student knowledge as a hidden state updated from each binary outcome. Recent deep learning versions like deep knowledge tracing (DKT) [64] and attentive knowledge tracing (AKT) [65] continue in this vein, using sequences of correct/incorrect answers to train neural networks that predict future performance. This binary-response context is appealingly easy to record—every learning management system (LMS) can log right or wrong answers—but it compresses a rich learning interaction into a single bit of information.

Recent research has been exploring richer representations of student responses. For example, question-centric knowledge tracing (qDKT) [66], option tracing [67], and open-ended knowledge tracing (OKT) [68] methods additionally incorporate question textual information, learners' actual selection in multiple choice questions, and learners' open-ended responses, respectively, into the classic KT framework, enabling both improved performance and additional insights. Regardless, it is striking that **much of today's learner modeling remains rooted in the** **same logs of question attempts and scores that have been collected for decades**, suggesting a vast untapped opportunity to widen the scope of captured context.

Beyond question–answer interactions, a wide range of additional contextual information may inform personalized instruction, if it can be captured reliably and ethically. One important category is learners' **affective and cognitive states**, such as engagement, confusion, or frustration, which are known to shape learning outcomes but are rarely observable in online systems. Recent work has explored multimodal approaches for inferring such states from classroom video or audio, for example, by combining facial expressions with surrounding visual context to improve academic emotion recognition [69]-[71] . While these approaches demonstrate the value of richer contextual signals, they also highlight practical challenges: continuous sensing can be intrusive, raises privacy concerns, and may not generalize across learning environments. As a result, a key open question is how to capture affective context in ways that are informative while being minimally disruptive and private for learners.

An additional increasingly important source of LC arises from chatbot-mediated learning interactions. As AI tutors and conversational agents become embedded in learning workflows [72], **chatbot logs capture rich traces of students' reasoning processes, misconceptions, and help-seeking behaviors**. Recent work has shown that such conversational data can be leveraged for learner modeling and knowledge tracing, for example by using LLMs to extract skill-level evidence from tutor–student dialogues [72]. Unlike traditional correctness logs, chatbot transcripts preserve *how* learners reason and respond over time. In parallel, industry systems are beginning to support persistent conversational context: modern AI assistants increasingly maintain memory and reusable skills across sessions (e.g., Claude's skills [73], OpenAI's memory feature [74], and other long-term memory mechanisms [75]), illustrating a broader shift toward infrastructure that supports durable, cross-session context accumulation. The education community is also moving towards such chatbot data infrastructure, such as the "tutoring observatory" that aims to collect large-scale tutoring dialogues [76] and data analytics frameworks for analyzing such tutoring conversations [77].

While the above advances build on context already captured within learning systems, emerging technologies also enable the **collection of richer activity-level context on learners' digital devices** with minimal interruption to learners' existing workflows. Recent work on GUI agents and operating-system-level instrumentation demonstrates the feasibility of capturing fine-grained digital activity traces—such as application switching, mouse and keyboard inputs, and screen recordings—across an entire device [78]-[82]. This is particularly important because **learning rarely occurs within a single application**: learners frequently move between digital textbooks, browsers, note-taking tools, and even physical media like pen and paper. Advances in automated annotation of screen recordings and interaction logs suggest new opportunities to reconstruct learners' trajectories across tools and tasks, yielding a more holistic view of learning behavior. However, realizing this potential will require substantial R&D on scalable data infrastructure, including methods to transform raw interaction traces into interpretable, privacy-aware LC representations.

### 3.3 Learning Context Prioritization and Compression

As the LC scales across learners, time, and institutions, its dimensionality will grow rapidly. Yet many contextual features contribute minimally to predictive or pedagogical power. A key R&D activity of Theme 2, therefore, involves a systematic study of which aspects of context matter most for learning outcomes and how to represent them efficiently.

**Prioritization framework.** Building on the theoretical weighting of LC dimensions in Theme 1, an important task is defining methods to determine the salience of each LC variable as a function of the current task and learner state. For example, affective indicators like students' motivation may receive higher weights during formative feedback, while social variables dominate in collaborative settings. Multi-objective optimization methods offer the potential to learn these weights by balancing predictive performance, interpretability, and privacy cost. Feature-importance analyses will identify which contextual dimensions most influence learning gains, enabling principled dimensionality reduction that remains faithful to theory.

**Compression and representation efficiency.** Three complementary strategies for LC compression hold promise:

- **Probabilistic feature selection:** Identifying minimal sufficient subsets of context variables using mutual-information and information-bottleneck criteria.
- **Context embedding reduction:** Training autoencoder-based models that map high-dimensional LC graphs to compact latent representations while preserving key relational properties.
- **Adaptive forgetting:** Introducing temporal decay and sparsification mechanisms that prune outdated or low-impact context information, consistent with cognitive theories of memory consolidation.

Compression performance must be evaluated not only for computational efficiency but also for pedagogical fidelity: how well the compressed LC preserves the theoretical meaning of the learner's cognitive and affective state. The resulting prioritized LC subgraph will serve as the working context that AI agents query in real time, while the full LC remains stored for longitudinal research analysis.

### 3.4 Learning Context Updating and Tracking

Learning is inherently dynamic; thus, the LC must evolve continually to reflect the learner's changing knowledge, goals, and environment. Guided by the theoretical "When" and "Where" dimensions of the LC framework, we must design temporal update mechanisms that treat the LC as a living model rather than a static record.

The LC must be updated as new evidence arrives to incorporate new information over outdated information as is typically done in time-series analysis. Previous work in KT (e.g., [66], [68]) will be crucial for modeling key components of learning dynamics. For example, the probability that a learner has mastered a concept may increase after consistent success but decay if later errors occur. Tracking must also address multi-platform learning: LC instances hosted by different learning systems should synchronize periodically through secure APIs, using hash-based verification, for example. Temporal smoothing and exponential decay functions will also be required to maintain contextual freshness without amplifying noise. Together, these methods will realize the theoretical principle that LC is a dynamic state variable that is constantly shaped by experience and interaction.

### 3.5 Optimizing Learning Interventions via the Learning Context

The ultimate measure of LC utility is its ability to improve teaching and learning decisions. The LC transforms from a passive data store into an active state representation that enables AI systems to reason about what instructional actions will most effectively advance learning.

**Context-aware decision models.** Reinforcement Learning (RL) approaches where the LC functions as the foundational state space can be used to **optimize pedagogical policies**. By utilizing the LC as a comprehensive "map" of the student's current state, the AI system can select the most effective instructional actions tailored to the individual, such as hint generation, problem sequencing, pacing adjustments, or modality shifts.

These models are designed to maximize **multi-objective rewards**, balancing raw learning gains with student engagement and affective stability. For example, as we saw in the shift from content-focused to learner-centered strategies for an anxious student in Vignette 2 above, the RL agent can employ the LC to prioritize emotional support alongside conceptual mastery.

As we have already discussed, the integration of "slow-to-surface" behavioral and linguistic traits into AI models is critical. Because the LC enables a "warm-start," the decision model does not have to spend the initial stages of a session "exploring" or guessing a learner's needs. Instead, it can immediately deploy a policy optimized for that student's specific behavioral or language proficiency levels from the first interaction.

Furthermore, context variables can serve a dual purpose: they can act both as predictive features to improve accuracy and as **ethical constraints** to ensure personalization remains pedagogically valid and equitable, preventing the AI from reinforcing biases based on a student's background.

**Evaluation and research questions.** Experimental studies within a variety of learning platforms will be needed to test whether LC-informed interventions outperform context-blind baselines. Primary research questions include: How much of the observed learning variance can LC-aware models explain beyond traditional models? Do context-sensitive interventions produce more durable learning? How do affective and social variables modulate optimal intervention timing?

Outcome measures should include learning efficiency (time-to-proficiency), engagement, changes in learning outcomes (e.g., pre-post tutoring session, course grades), start persistence, affective stability, and instructor interpretability. These studies will complete the theory-to-practice loop: hypotheses derived from LC theory are implemented algorithmically and evaluated empirically, advancing both learning science and AI pedagogy as we produce a **richer personalization of AI tools** (Figure 5, bottom right)**.**

### 3.6 Learning Context Implementation via the Model Context Protocol (MCP)

Given the LC theoretical framework and its computational implementation, the final challenge is interfacing this data with diverse AI tools. Fortunately, recent developments in industry standards substantially streamline this integration. The Model Context Protocol (MCP)—introduced by Anthropic in late 2024 and now supported by major providers like OpenAI and Google—provides the open, standardized framework necessary for AI systems to securely exchange data with external resources. MCP provides the "plug-and-play" capability that turns the LC from a static database into a portable educational passport.

**MCP as the "warm-start" catalyst.** While current AI tutors typically start every session "context-blind," MCP can act as the integration layer that enables these tools to query the LC server immediately upon initialization (Figure 5, bottom right).

- **Eliminating the cold-start:** Instead of a learner spending hours "re-proving" their language proficiency or behavioral traits, an AI tool can use MCP to issue a structured query: "What are the key cognitive misconceptions and affective traits for this learner?"
- **Durable signal preservation:** Because recovery of behavioral traits like conscientiousness is slow (often taking multiple interactions; recall Vignette 4 above), MCP can ensure that these signals are preserved and "travel" with the student across tools.
- **Bidirectional synchronization:** MCP supports secure, bidirectional communication, enabling AI models to both pull from the LC to personalize instruction and push new evidence back to the LC to update the learner's profile in real-time.

**Privacy-first interoperability.** By adopting the MCP standard, we can decouple the AI system's "reasoning" from its raw "data storage". We discuss further below in Section 5.

- **Data minimization:** AI models receive structured snapshots of context without requiring direct access to raw, personally identifiable information (PII).
- **Reduced technical friction:** Standardized metadata ensures that the LC can interoperate broadly across different educational platforms with minimal custom code.
- **Governance and compliance:** This architecture enables centralized control over data retention and consent, ensuring that LC innovations remain compliant with ethical and privacy requirements while maximizing instructional utility.

## 4. Theme 3: Application, Testing, and Validation

We view the "arrival" of AI as an opportunity to build a more **personalized, equitable, and interpretable** learning ecosystem. We posit that each learner's **context** (e.g., prior knowledge, cultural background, affect, strategies, and accommodations) is an asset that should travel with the learner across tools. Today's AI tools largely operate **context-blind**, producing fragmented insight across LMSs, homework systems, and discussion spaces. We are working to deliver the **infrastructure and methods** needed to (i) **represent LC** in human-interpretable, machine-actionable form; (ii) **port** that context across AI platforms so supports follow the learner; and (iii) establish **governance** (consent, minimization, provenance, retention) that others can adopt. We anticipate downstream benefits in persistence, better instructional decisions, and cost-effective personalization for resource-constrained institutions.

### 4.1 Problems of Practice in Authentic Educational Settings

Across the board in education, processes abound where context information can lend insight into the implementation of educational tools and resources in the classroom. Multiple systems build partial learner models that do not share meaning. Consequences include: AI that cannot follow learners across modalities, instructors who cannot interpret feedback across tools, and learners who receive **inconsistent guidance** as they move between digital and in-person learning. We briefly review our current work towards answering the foundational research question: Does context-aware AI tutoring outperform context-blind AI tutoring in enhancing learner engagement, persistence, and learning outcomes in authentic educational settings?

OpenStax [83] offers a national, LMS-integrated ecosystem in which this fragmentation exists and is addressable. With **85+ open textbooks** and the **Assignable** teaching and learning platform [12] used at scale, OpenStax provides the authentic setting for integrating, testing, and evaluating context-aware AI. **SafeInsights** [13], a privacy-preserving research infrastructure housed at Rice University, enables analysis **where the data live** (institutional or OpenStax environments) so that LMS, Assignable, and optional self-report/context sources can be combined **responsibly** without exporting raw PII. **OpenStax Kinetic** [4] provides survey delivery and a Learner Characteristics Library to enrich context with validated constructs. Partnerships with leading AI platforms ensure interoperability and real-world portability testing across multiple tools educators already use.

OpenStax's growing network of partnerships with leading AI and edtech organizations provides a powerful ecosystem for advancing context-aware learning research. Collaborations with Google [85] and Microsoft Education [86], among others, extend these capabilities to generative tutoring, symbolic reasoning, and interactive instructional design. Together, these collaborations are enabling us to embed, test, and refine the context-representation framework within multiple authentic educational AI environments, ensuring interoperability, ethical data sharing, and scalability based on the very tools educators and learners already use.

### 4.2 An Integrated Co-Design Methodology: Blending Design Based Implementation Research and User-Centered Methodology

Our work blends design-based implementation research (DBIR) [87], [88] for our research-practice partnership. with user-centered design (UCD) [84] to ensure solutions are usable, trusted, and sustainable.

**DBIR.** With instructors and learners, we jointly define the context constructs most likely to improve teaching/learning, identify risks and mitigations, and iteratively refine implementation in real courses. We collect context-rich evidence (surveys, interviews, observations, usage) to inform both product and practice. We build community capacity by sharing tools, templates, and examples.

**UCD.** Guided by Human Computer Interaction (HCI) and User Experience (UX) expertise, we run an agile design lifecycle (needs analysis, prototyping, heuristic and task evaluations, SUS/utility checks). Prototypes flow back to DBIR partners to assess pedagogical fit and feasibility. This continuous loop is essential for trustworthy AI, aligning assistive behavior, explanations, and controls with instructor goals and classroom realities.

A critical feature of this integrated model is the **continuous feedback loop** between the frameworks. UCD-driven prototypes will be tested for usability and utility (e.g., using the System Usability Scale (SUS) [89]. These findings will then be brought back to the DBIR partnership to assess pedagogical fit, implementation feasibility, and potential barriers to adoption. This is particularly vital for AI-driven tools, where co-design is essential for building educator trust, ensuring pedagogical alignment, and managing the complexities of classroom integration [90]. In addition to leveraging the Kinetic Learner Characteristics Library, we will use various instruments to measure learning LC while examining interactions with OpenStax Assignable and the AI tools, closing the formative loop that informs both product refinement (UCD) and implementation strategy (DBIR).

### 4.3 Research Design and Methods

We are currently conducting a pilot study to investigate how context affects instructional adjustments and personalization in AI tools. For examining the effect of context-aware AI tools for instruction in real-world courses and classrooms [91], we are utilizing OpenStax Assignable integrated with SafeInsights, a national, privacy-first R&D research infrastructure. The context information will be stored in Assignable with provenance (source, timestamps), consent, and retention information.

We are using within-course factorials (context-aware AI tutoring vs. context-blind AI tutoring vs. Standard Practice); and micro-randomized trials [92] (randomize when context is injected) to estimate the immediate causal "footprint" of context on next-item success. We are also using the SafeInsights privacy-preserving research infrastructure for all in-course studies, ensuring analyses occur where the data live (institutional or OpenStax environments) and that only vetted aggregates leave the enclave.

### 4.4 Evaluation and Success Criteria

To apply and evaluate the LC implementation laid out in Theme 2, we are actively working to contrast AI tutoring experiences when the tool is **context-aware** (tool consumes the LC via the context server/MCP) vs. when it is **context-agnostic** (same tool, no external context; only the current prompt/task). We validate the context pipeline along three layers: technical, semantic, and functional, then monitor reliability, fairness, and privacy in situ.

**Technical validity.** We verify schema conformance, completeness, timestamp integrity, PII separation, and ensure all records have auditable metadata (provenance, consent, retention tags).

**Semantic validity (Does a context feature mean what we think?).** We map context features to learning-science constructs [93] and validate them against independent measures (e.g., graded work) and human expert review.

**Functional validity (Is the context used correctly?).** We use context-fidelity probes (embedded test cases) to ensure the tool uses context correctly, respects restrictions, and handles missing data gracefully. We use micro-randomized trials to estimate the causal impact of the context.

**Reliability, portability, and drift.** We check temporal stability, ensure consistent behavior across different tools, and run scheduled checks for model drift during deployment.

**Benchmark layer for AI tutoring system.** Before we attribute gains to context awareness, we need to (i) control for raw domain mastery, (ii) verify core tutoring moves (explanations, hints, feedback), (iii) check dialogic/Socratic skill, and (iv) test whether models can update from language feedback, the mechanism we use to refine the learner's context graph. We will use established benchmarks such as Humanity's Last Exam (HLE) [94], TutorBench [95], SocraticBench [96], and LLF Bench [97], respectively, to set baselines for each of the four dimensions.

**Success criteria.** We define clear "go/no-go" criteria for success. The full context-aware system will only be considered for broader deployment if it meets the following minimum thresholds in our efficacy trials: (i) ***Efficacy.*** The context-aware system must produce learning gains significantly larger than the context-blind control, achieving a minimum effect size of **Cohen's *d*** ≥ 0.20 on primary learning outcomes. (ii) ***Fidelity and safety.*** The context model must demonstrate high predictive validity (e.g., ≥ 95% accuracy in classifying learner states against ground-truth measures) and commit zero critical data-privacy or scope violations. (iii) ***Practicality.*** The system's computational costs (e.g., latency and server load) must be low enough to function robustly within the technical and budgetary constraints of a typical educational setting. If these thresholds are not met, our methodology includes a clear "iterate-and-retest" loop, where we will refine context features and compression models before a new validation test.

## 5. Theme 4: Privacy and Ethics

While the LC framework offers transformative potential for personalized, context-aware AI teaching and learning, it also introduces **significant challenges related to privacy, security, governance, and the ethical use of personal information**, which varies in terms of identifiability [110]. This theme cuts across all others, ensuring that privacy and ethics are intentionally embedded in each phase of LC design, implementation, and evaluation rather than treated as an afterthought. Given the pervasive use of AI in high-stakes applications, now more than ever, individuals need to understand and have agency over how their data is being used [98].

A key guiding principle for LC R&D is that privacy and learning performance exist in tension: as privacy protections are strengthened (by an amount $\varepsilon$), system utility and learning outcomes often degrade (by $\delta$). Our recommended approach, therefore, proceeds in two stages. First, we suggest establishing the upper bound of system performance: **how effectively can an LC-informed AI system guide learning when unconstrained by privacy restrictions?** Second, we suggest integrating privacy-preserving mechanisms and empirically quantifying their impact, balancing the tradeoff between privacy ($\varepsilon$) and performance ($\delta$). This approach will yield a principled understanding of where, and how, privacy constraints most strongly affect learning effectiveness. Our research will be guided by four central questions:

**<u>Governance and ownership: Who should own and control the learning context</u>?** With the OpenStax learner and educator communities, for example, we can investigate governance models across educational settings, such as different educational levels, environments, and state and federal legal lines. These models should be evaluated in terms of transparency, consent, portability, and accountability [99].

**<u>Secure architectural tradeoffs</u>.** To determine the optimal architecture for storing and safeguarding sensitive context data, we suggest examining centralized, decentralized [100], federated [101], and distributed [102] (including blockchain-linked [103]) models, exploring hybrid architectures in which fragments of the LC are distributed across multiple enclaves or services. A distributed approach can reduce risk by making it computationally infeasible for any single breach to reconstruct a full learner profile. Crucially, how can we safeguard against "inference attacks" where AI models might reconstruct or leak sensitive attributes? We will investigate frameworks that treat the AI model as an "untrusted" agent [109], validating that even within secure enclaves, model outputs remain confidential and do not inadvertently expose protected data.

**<u>Controlled disclosure mechanisms</u>.** How can AI tools benefit from learner context without direct access to raw PII? We are exploring privacy-preserving computation [104] and data minimization strategies, including query-based access patterns, differential privacy budgets, and adaptive "blurring" techniques that tune the granularity of shared information to match instructional intent [105].

**Contextual fairness and ethical alignment.** How do we ensure the LC acts as a support rather than a limitation? There is a risk that AI models might use a student's background or past performance to lower expectations, effectively reinforcing bias [41]. We suggest investigating methods to detect and prevent these patterns, ensuring that personalization maintains high academic standards and expands opportunities for all learners, rather than restricting a student's future based on their LC.

To ensure responsible design, LC R&D should be grounded in a blended DBIR-UCD process that directly involves learners and instructors. These stakeholders will help define acceptable privacy/performance tradeoffs and provide input on transparency, consent, and interpretability. Our design philosophy rejects systems that unilaterally define a learner's context. Instead, the LC must be built to enhance user agency, offering interpretable representations for instructors and transparent controls for learners.

Finally, we emphasize that **ethics in AI for education extends beyond compliance**. Ethical integration requires a commitment to fairness, accountability, transparency, and explainability. It is crucial that the emerging LC community documents and publishes ethical design guidelines, informed by their empirical results, for broader dissemination across the AI-in-education community. Through this cross-cutting effort, we can pioneer a model for responsible innovation in AI-powered learning that safeguards privacy while advancing the scientific understanding of how context can enhance learning outcomes.

## 6. Conclusions

While Generative AI has triggered a seismic change in how information is accessed, today's AI tools remain mostly context-blind. They are powerful mimics but fundamentally incapable of truly understanding the individual learner. We have proposed a principled transition away from fragmented, tool-isolated interactions and towards a unified LC that grows and evolves alongside the student.

**From warm-starts to evolving intelligence.** The LC framework does more than solve the cold-start problem; it establishes a foundation for continual, long-term personalization.

- **A living model:** The LC is designed as a dynamic data structure, not a static record. It evolves continually to reflect a learner's changing knowledge, goals, and environment.
- **Capturing the "whole" learner:** By formalizing cognitive, affective, and sociocultural factors, AI systems leveraging the LC will be able to better understand who they are teaching and not just what content to deliver.
- **Cross-platform growth:** Because learning rarely occurs within a single application, the LC is built to travel with the learner across different AI systems, textbooks, browsers, and note-taking tools.

**A pathway for global impact.** Our recommendations and explorations with OpenStax and SafeInsights are not merely a theoretical exercise but a practical, scalable roadmap to catalyze safer and more equitable AI-based teaching and learning nationwide.

- **Broad implementation:** The LC framework is designed for broad adoption across a wide range of digital education platforms. By emphasizing open standards, current and emergent edtech providers can leverage these data structures to fundamentally improve learning outcomes at scale.
- **Privacy-first implementation and validation:** We have strongly recommended a privacy-by-design approach, where protections are intentionally embedded into every phase of LC design and implementation.
- **Closing equity gaps:** Context-aware AI can promote deep, robust learning rather than shallow mimicry, ensuring that this technology minimizes equity gaps instead of widening them.

**The human-technology frontier.** Ultimately, the LC framework can enable a human-AI partnership where the AI model learns with the learner over time. By helping AI systems perceive, reason about, and adapt to the evolving activities of students, we can move toward a future where education is not one-size-fits-all, but is as unique and dynamic as the individuals it serves. We invite the R&D community to join us in realizing this shift from context-blind tools to context-aware learning companions.

**Contact:** Richard Baraniuk, richb@rice.edu

Acknowledgements: This work was supported by NSF SafeInsights 2153481 and ONR MURI N00014-20-1-2787.

## References

[1] J. Reich and J. Dukes, "Toward a new theory of arrival technologies," *EdArXiv Preprints*, 2024.

[2] A. Chatterji *et al.*, "How people use ChatGPT," National Bureau of Economic Research, Working Paper, Sep. 2025.

[3] S. C. Kong and Y. Yang, "A human-centered learning and teaching framework using generative artificial intelligence for self-regulated learning development through domain knowledge learning in K–12 settings," *IEEE Trans. Learn. Technol.*, vol. 17, pp. 1562–1573, Apr. 2024.

[4] D. Basu Mallick, B. C. Bradford, and R. Baraniuk, "Secure education and learning research at scale with OpenStax Kinetic," in *Proc. 10th ACM Conf. Learning @ Scale*, Jul. 2023, pp. 360–362.

[5] J. H. Kaufman, A. Woo, J. Eagan, S. Lee, and E. B. Kassan, "Uneven adoption of artificial intelligence tools among US teachers and principals in the 2023-2024 school year," RAND Corporation, Santa Monica, CA, Rep. RR-A123-1, 2025.

[6] K. Karpouzis, D. Pantazatos, J. Taouki, and K. Meli, "Tailoring education with GenAI: A new horizon in lesson planning," in *2024 IEEE Global Eng. Educ. Conf. (EDUCON)*, May 2024, pp. 1–10.

[7] E. Kasneci *et al.*, "ChatGPT for good? On opportunities and challenges of large language models for education," *Learn. Individ. Differ.*, vol. 103, Art. no. 102274, Apr. 2023.

[8] X. Hou, Y. Zhao, S. Wang, and H. Wang, "Model context protocol (MCP): Landscape, security threats, and future research directions," *arXiv preprint arXiv:2503.23278*, Mar. 2025.

[9] K. Milam and A. Gulli, "Context engineering: Sessions & memory," *Kaggle*, 2025. [Online].

[10] D. Osher, P. Cantor, J. Berg, L. Steyer, and T. Rose, "Drivers of human development: How relationships and context shape learning and development," *Appl. Developmental Sci.*, vol. 24, no. 1, pp. 6–36, Jan. 2020.

[11] P. Cantor, D. Osher, J. Berg, L. Steyer, and T. Rose, "Malleability, plasticity, and individuality: How children learn and develop in context," in *The Science of Learning and Development*. New York, NY, USA: Routledge, 2021, pp. 3–54.

[12] OpenStax, "Assignable," *OpenStax*. [Online]. Available: https://openstax.org/assignable/

[13] Safe Insights, "Safe Insights," *Safe Insights*. [Online]. Available: https://safeinsights.org

[14] Digital Promise, "Learner Variability Project," *DigitalPromise.org*, 2025. [Online]. Available: https://digitalpromise.org/initiative/learner-variability-project/

[15] P. R. Pintrich, D. A. Smith, T. Garcia, and W. J. McKeachie, "A manual for the use of the Motivated Strategies for Learning Questionnaire (MSLQ)," Nat. Center for Research to Improve Postsecondary Teaching and Learning, Ann Arbor, MI, 1991.

[16] N. Selwyn, "On the limits of artificial intelligence (AI) in education," *Nordisk tidsskrift for pedagogikk og kritikk*, vol. 10, no. 1, pp. 3–14, 2024.

[17] National Academies of Sciences, Engineering, and Medicine, *How People Learn II: Learners, Contexts, and Cultures*. Washington, DC, USA: National Academies Press, 2018.

[18] B. Barron, "Interest and self-sustained learning as catalysts of development: A learning ecology perspective," *Hum. Dev.*, vol. 49, no. 4, pp. 193–224, Oct. 2006.

[19] L. S. Vygotsky, *Mind in Society: The Development of Higher Psychological Processes*. Cambridge, MA, USA: Harvard Univ. Press, 1978.

[20] J. S. Brown, A. Collins, and P. Duguid, "Situated cognition and the culture of learning," *Educ. Researcher*, vol. 18, no. 1, pp. 32–42, Feb. 1989.

[21] J. Lave and E. Wenger, *Situated Learning: Legitimate Peripheral Participation*. Cambridge, U.K.: Cambridge Univ. Press, 1991.

[22] C. T. Fosnot, *Constructivism: Theory, Perspectives, and Practice*. New York, NY, USA: Teachers College Press, 2013.

[23] E. Hutchins, *Cognition in the Wild*. Cambridge, MA, USA: MIT Press, 1995.

[24] J. Hollan, E. Hutchins, and D. Kirsh, "Distributed cognition: Toward a new foundation for human–computer interaction research," *ACM Trans. Comput.-Hum. Interact.*, vol. 7, no. 2, pp. 174–196, 2000.

[25] U. Bronfenbrenner and S. J. Ceci, "Nature-nurture reconceptualized in developmental perspective: A bioecological model," *Psychol. Rev.*, vol. 101, no. 4, p. 568, 1994.

[26] T. Burner and B. Svendsen, "Activity theory—Lev Vygotsky, Aleksei Leont'ev, Yrjö Engeström," in *Springer International Handbooks of Education*. Cham, Switzerland: Springer, 2025, pp. 313–327.

[27] B. Rogoff, *Apprenticeship in Thinking: Cognitive Development in Social Context*. New York, NY, USA: Oxford Univ. Press, 1990.

[28] J. V. Wertsch, *Voices of the Mind: A Sociocultural Approach to Mediated Action*. Hemel Hempstead, U.K.: Harvester Wheatsheaf, 1991.

[29] D. A. Kolb, *Experiential Learning: Experience as the Source of Learning and Development*. Englewood Cliffs, NJ, USA: Prentice Hall, 1984.

[30] D. H. Jonassen and L. Rohrer-Murphy, "Activity theory as a framework for designing constructivist learning environments," *Educ. Technol. Res. Dev.*, vol. 47, no. 1, pp. 61–79, Mar. 1999.

[31] J. Roschelle, E. A. McLaughlin, and K. R. Koedinger, "Beyond benchmarks: Responsible AI in education needs learning sciences," *Commun. ACM*, Jan. 2025.

[32] K. VanLehn, "The relative effectiveness of human tutoring, intelligent tutoring systems, and other tutoring systems," *Educ. Psychol.*, vol. 46, no. 4, pp. 197–221, Oct. 2011.

[33] P. L. Ackerman, M. E. Beier, and M. O. Boyle, "Working memory and intelligence: The same or different constructs?" *Psychol. Bull.*, vol. 131, no. 1, p. 30, Jan. 2005.

[34] M. Schneider and B. A. Simonsmeier, “How does prior knowledge affect learning? A review of 16 mechanisms and a framework for future research,” *Learn. Individ. Differ.*, vol. 122, Art. no. 102744, Aug. 2025.

[35] A. E. Poropat, “A meta-analysis of the five-factor model of personality and academic performance,” *Psychol. Bull.*, vol. 135, no. 2, p. 322, Mar.

[36] X. Huang and R. E. Mayer, “Benefits of adding anxiety-reducing features to a computer-based multimedia lesson on statistics,” *Comput. Hum. Behav.*, vol. 63, pp. 293–303, Oct. 2016.

[37] R. W. Engle, “Working memory capacity as executive attention,” *Curr. Dir. Psychol. Sci.*, vol. 11, no. 1, pp. 19–23, Feb. 2002.

[38] J. Shin, R. Balyan, M. P. Banawan, T. Arner, W. L. Leite, and D. S. McNamara, “Analyzing interaction patterns and content dynamics in an online mathematics discussion board,” *Interact. Learn. Environ.*, vol. 33, no. 3, pp. 2151–2174, Mar. 2025.

[39] J. G. Pyke and J. J. Sherlock, “A closer look at instructor-student feedback online: A case study analysis of the types and frequency,” *MERLOT J. Online Learn. Teach.*, vol. 6, no. 1, pp. 110–121, Mar. 2010.

[40] J. Xie and A.-P. Correia, “The effects of instructor participation in asynchronous online discussions on student performance: A systematic review,” *Brit. J. Educ. Technol.*, vol. 55, no. 1, pp. 71–89, Jan. 2024.

[41] R. S. Baker and A. Hawn, “Algorithmic bias in education,” *Int. J. Artif. Intell. Educ.*, vol. 32, no. 4, pp. 1052–1092, 2022.

[42] P. Robinson, “Task complexity, task difficulty, and task production: Exploring interactions in a componential framework,” *Appl. Linguist.*, vol. 22, no. 1, pp. 27–57, Mar. 2001.

[43] J. Rho, M. A. Rau, and B. D. Van Veen, “Preparing collaborative future learning with representational-competency supports,” in *Proc. 17th Int. Conf. Learn. Sci. (ICLS)*, 2023, pp. 43–50.

[44] G. Mazzeo Ortolani, “STEM and STEAM education, and disciplinary integration: A guide to informed policy action,” European Commission, Brussels, Belgium, Rep. JRC141438, 2025.

[45] S. Park, “Analysis of time-on-task, behavior experiences, and performance in two online courses with different authentic learning tasks,” *Int. Rev. Res. Open Distrib. Learn.*, vol. 18, no. 2, pp. 213–233, 2017.

[46] T. Tchoubar, T. R. Sexton, and L. L. Scarlatos, “Role of digital fluency and spatial ability in student experience of online learning environments: Digital readiness for evolution of educational ecosystem,” in *Proc. Science and Information Conf.*, 2018, pp. 524–532.

[47] J. Reeve and W. Lee, “Students’ classroom engagement produces longitudinal changes in classroom motivation,” *J. Educ. Psychol.*, vol. 106, no. 2, pp. 527–540, May 2014.

[48] R. Baker, W. Ma, Y. Zhao, S. Wang, and Z. Ma, “The results of implementing Zone of Proximal Development on learning outcomes,” in *Proc. 13th Int. Conf. Educ. Data Mining (EDM)*, Jul. 2020, pp. 749–753.

[49] W. Crain, *Theories of Development: Concepts and Applications*, 6th ed. New York, NY, USA: Psychology Press, 2015.

[50] K. R. Koedinger, J. L. Booth, and D. Klahr, "Instructional complexity and the science needed to constrain it," *Science*, vol. 342, no. 6161, pp. 935–937, Nov. 2013.

[51] S. Nawaz, N. Srivastava, J. H. Yu, R. S. Baker, G. Kennedy, and J. Bailey, "Analysis of task difficulty sequences in a simulation-based POE environment," in *Artif. Intell. Educ. (AIED)*, 2020, pp. 423–436.

[52] M. L. Bernacki and C. A. Walkington, "The impact of a personalization intervention for mathematics on learning and non-cognitive factors," in *Proc. Educ. Data Mining (EDM) Workshops*, 2014.

[53] A. Grubišić, S. Stankov, and B. Žitko, "Adaptive courseware: A literature review," *J. Univers. Comput. Sci.*, vol. 21, no. 9, pp. 1168–1209, 2015.

[54] G. D. Rey, "A review of research and a meta-analysis of the seductive detail effect," *Educ. Res. Rev.*, vol. 7, no. 3, pp. 216–237, Dec. 2012.

[55] R. Mayer, "Ten research-based principles for designing multimedia instruction," in *Proc. E-Learn: World Conf. E-Learning*, 2014, pp. 1262–1273.

[56] R. D. Roscoe and D. S. McNamara, "Writing Pal: Feasibility of an intelligent writing strategy tutor in the high school classroom," *J. Educ. Psychol.*, vol. 105, no. 4, pp. 1010–1025, Nov. 2013.

[57] J. Wong, M. Baars, D. Davis, T. Van Der Zee, G.-J. Houben, and F. Paas, "Supporting self-regulated learning in online learning environments and MOOCs: A systematic review," *Int. J. Hum.-Comput. Interact.*, vol. 35, no. 4–5, pp. 356–373, 2019.

[58] Advanced Distributed Learning Initiative, "Experience API (xAPI) specification, version 1.0.3," ADL, Alexandria, VA, USA, 2016. [Online]. Available: https://github.com/adlnet/xAPI-Spec

[59] W3C, "JSON-LD 1.1: A JSON-based serialization for linked data," World Wide Web Consortium, Rec., Feb. 2020. [Online]. Available: https://www.w3.org/TR/json-ld11/

[60] Neo4j Inc., "Neo4j graph database platform." [Online]. Available: https://neo4j.com/

[61] S. Sonkar, A. Katiyar, and R. Baraniuk, "NePTuNe: Neural powered Tucker network for knowledge graph completion," in *Proc. 10th Int. Joint Conf. Knowl. Graphs (IJCKG)*, Dec. 2021, pp. 177–180.

[62] J. Hatchett, D. B. Mallick, and R. Baraniuk, "From words to wisdom: Automatically generating knowledge graphs for interpretable educational AI," in *Proc. 38th AAAI Annu. Conf. Artif. Intell. (AAAI)*, 2024.

[63] A. T. Corbett and J. R. Anderson, "Knowledge tracing: Modeling the acquisition of procedural knowledge," *User Model. User-Adapt. Interact.*, vol. 4, no. 4, pp. 253–278, 1994.

[64] C. Piech *et al.*, "Deep knowledge tracing," *arXiv preprint arXiv:1506.05908*, 2015.

[65] A. Ghosh, N. Heffernan, and A. S. Lan, "Context-aware attentive knowledge tracing," *arXiv preprint arXiv:2007.12324*, 2020.

[66] S. Sonkar, A. E. Waters, A. S. Lan, P. J. Grimaldi, and R. G. Baraniuk, "qDKT: Question-centric deep knowledge tracing," in *Proc. 13th Int. Conf. Educ. Data Mining (EDM)*, 2020, pp. 677–681.

[67] A. Ghosh, J. Raspat, and A. Lan, “Option tracing: Beyond correctness analysis in knowledge tracing,” *arXiv preprint arXiv:2104.09043*, Apr. 2021.

[68] N. Liu, Z. Wang, R. Baraniuk, and A. Lan, “Open-ended knowledge tracing for computer science education,” in *Proc. 2022 Conf. Empir. Methods Nat. Lang. Process. (EMNLP)*, 2022.

[69] L. Zhao, J. Xuan, J. Lou, Y. Yu, and W. Yang, “Context-aware academic emotion dataset and benchmark,” *arXiv preprint arXiv:2507.00586*, 2025.

[70] N. Bosch *et al.*, “Detecting student emotions in computer-enabled classrooms,” in *Proc. 25th Int. Joint Conf. Artif. Intell. (IJCAI)*, 2016, pp. 4125–4129.

[71] A. S. Lan, A. F. Botelho, S. Karumbaiah, R. S. Baker, and N. Heffernan, “Accurate and interpretable sensor-free affect detectors via monotonic neural networks,” in *Companion Proc. 10th Int. Conf. Learn. Analytics & Knowl. (LAK)*, 2020.

[72] A. Scarlatos, R. S. Baker, and A. S. Lan, “Exploring knowledge tracing in tutor-student dialogues using LLMs,” *arXiv preprint arXiv:2409.16490*, 2024.

[73] Anthropic, “Agent skills — Claude code documentation,” *Claude Code Docs*, 2025. [Online]. Available: https://code.claude.com/docs/en/skills

[74] OpenAI, “Memory and new controls for ChatGPT,” Feb. 13, 2024. [Online]. Available: https://openai.com/index/memory-and-new-controls-for-chatgpt/

[75] W. Xu, Z. Liang, K. Mei, H. Gao, J. Tan, and Y. Zhang, “A-MEM: Agentic memory for LLM agents,” *arXiv preprint arXiv:2502.12110*, 2025.

[76] D. R. Thomas *et al.*, “Advancing the science of teaching with tutoring data: A collaborative workshop with the National Tutoring Observatory,” in *Proc. 12th ACM Conf. Learning @ Scale (L@S)*, 2025.

[77] R. E. Wang and D. Demszky, “Edu-ConvoKit: An open-source library for education conversation data,” *arXiv preprint arXiv:2402.05111*, 2024.

[78] T. Xie *et al.*, “OSWorld: Benchmarking multimodal agents for open-ended tasks in real computer environments,” *arXiv preprint arXiv:2404.07972*, 2024.

[79] Z. Z. Wang, Y. Shao, O. Shaikh, D. Fried, G. Neubig, and D. Yang, “How do AI agents do human work? Comparing AI and human workflows across diverse occupations,” *arXiv preprint arXiv:2510.22780*, 2025.

[80] P. Yang, H. Ci, and M. Z. Shou, “macOSWorld: A multilingual interactive benchmark for GUI agents,” *arXiv preprint arXiv:2506.04135*, 2025.

[81] X. Wang *et al.*, “OpenCUA: Open foundations for computer-use agents,” *arXiv preprint arXiv:2508.09123*, 2025.

[82] J. Chen *et al.*, “SPA-Bench: A comprehensive benchmark for smartphone agent evaluation,” arXiv preprint arXiv:2410.15164, 2024.

[83] OpenStax, “OpenStax,” Rice University. [Online]. Available: https://openstax.org

[84] R. D. Pea, “User centered system design: New perspectives on human-computer interaction,” *J. Educ. Comput. Res.*, vol. 3, no. 1, pp. 129–134, 1987.

[85] The OpenStax Team, “Press release: OpenStax partners with Google’s Gemini apps,” OpenStax.org, Aug. 29, 2024. [Online]. Available: https://openstax.org/blog/press-release-openstax-partners-google-gemini-apps

[86] OpenStax, “OpenStax and Microsoft partner for AI-enhanced learning in the classroom,” OpenStax Blog, 2024. [Online]. Available: https://openstax.org/blog/press-release-openstax-and-microsoft-partner-for-ai-enhanced-learning-in-the-classroom

[87] S. M. Underwood and A. T. Kararo, “Design-based implementation research (DBIR): An approach to propagate a transformed general chemistry curriculum across multiple institutions,” *J. Chem. Educ.*, vol. 98, no. 12, pp. 3643–3655, Dec. 2021.

[88] B. J. Fishman, W. R. Penuel, A.-R. Allen, B. H. Cheng, and N. Sabelli, “Design-based implementation research: An emerging model for transforming the relationship of research and practice,” *Teach. Coll. Rec.*, vol. 115, no. 14, pp. 1–28, 2013.

[89] J. Brooke, “SUS: A ‘quick and dirty’ usability scale,” in *Usability Evaluation in Industry*, P. W. Jordan, B. Thomas, B. A. Weerdmeester, and I. L. McClelland, Eds. London, U.K.: Taylor & Francis, 1996, pp. 189–194.

[90] R. A. Grier, A. Bangor, P. Kortum, and S. C. Peres, “The system usability scale: Beyond standard usability testing,” in *Proc. Hum. Factors Ergonom. Soc. Annu. Meeting*, Los Angeles, CA, USA, 2013, pp. 187–191.

[91] K. Holstein, B. M. McLaren, and V. Aleven, “Co-designing a real-time classroom orchestration tool to support teacher-AI complementation,” in *Proc. 9th Int. Conf. Learn. Anal. Knowl. (LAK)*, Tempe, AZ, USA, Mar. 2019, pp. 299–308.

[92] U.S. National Science Foundation, “NSF invests $90M in innovative national scientific cyberinfrastructure for transforming STEM education,” NSF.gov, Apr. 24, 2024. [Online]. Available: https://new.nsf.gov/news/nsf-invests-90m-innovative-national-scientific

[93] P. Klasnja *et al.*, “Microrandomized trials: An experimental design for developing just-in-time adaptive interventions,” *Health Psychol.*, vol. 34, no. S, pp. 1220–1228, Dec. 2015.

[94] L. J. Cronbach and P. E. Meehl, “Construct validity in psychological tests,” *Psychol. Bull.*, vol. 52, no. 4, pp. 281–302, 1955.

[95] L. Phan *et al.*, “Humanity’s last exam,” arXiv preprint arXiv:2501.14249, Jan. 2025.

[96] R. S. Srinivasa *et al.*, “TutorBench: A benchmark to assess tutoring capabilities of large language models,” arXiv preprint arXiv:2510.02663, Oct. 2025.

[97] G. Gatti, “Socratic-Bench,” GitHub, 2023. [Online]. Available: https://github.com/GiovanniGatti/socratic-bench

[98] “How A.I. can use your personal data to hurt your neighbor,” *The New York Times*, Nov. 2, 2025. [Online]. Available: https://www.nytimes.com/2025/11/02/opinion/ai-privacy.html

[99] K. Finch, “A visual guide to practical data de-identification,” *Future of Privacy Forum*, Apr. 25, 2016. [Online]. Available: https://fpf.org/blog/a-visual-guide-to-practical-data-de-identification/

[100] P. Maymounkov and D. Mazieres, “Kademlia: A peer-to-peer information system based on the XOR metric,” in *Proc. 1st Int. Workshop Peer-to-Peer Syst. (IPTPS)*, Cambridge, MA, USA, Mar. 2002, pp. 53–65.

[101] B. McMahan, E. Moore, D. Ramage, S. Hampson, and B. A. y Arcas, “Communication-efficient learning of deep networks from decentralized data,” in *Proc. 20th Int. Conf. Artif. Intell. Stat. (AISTATS)*, Fort Lauderdale, FL, USA, Apr. 2017, pp. 1273–1282.

[102] J. Verbraeken *et al.*, “A survey on distributed machine learning,” *ACM Comput. Surv.*, vol. 53, no. 2, pp. 1–33, Mar. 2020.

[103] G. Chen, B. Xu, M. Lu, and N.-S. Chen, “Exploring blockchain technology and its potential applications for education,” *Smart Learn. Environ.*, vol. 5, no. 1, p. 1, Dec. 2018.

[104] C. Dwork and A. Roth, “The algorithmic foundations of differential privacy,” *Found. Trends Theor. Comput. Sci.*, vol. 9, no. 3–4, pp. 211–407, Aug. 2014.

[105] A. Acar, H. Aksu, A. S. Uluagac, and M. Conti, “A survey on homomorphic encryption schemes: Theory and implementation,” *ACM Comput. Surv.*, vol. 51, no. 4, pp. 1–35, Jul. 2019.

[106] S. Sonkar, X. Chen, N. Liu, R. G. Baraniuk, and M. Sachan, “LLM-based cognitive models of students with misconceptions,” *arXiv preprint arXiv:2410.12294*, Oct. 2024.

[107] S. Sonkar, N. Liu, and R. Baraniuk, “Student data paradox and curious case of single student-tutor model: Regressive side effects of training LLMs for personalized learning,” in *Findings of the Assoc. for Comput. Linguist.: EMNLP 2024*, Nov. 2024, pp. 15543–15553.

[108] S. Sonkar, K. Ni, S. Chaudhary, and R. Baraniuk, “Pedagogical alignment of large language models,” in *Findings of the Assoc. for Comput. Linguist.: EMNLP 2024*, Nov. 2024, pp. 13641–13650.

[109] F. Tramer and D. Boneh, “Slalom: Fast, verifiable and private execution of neural networks in trusted hardware,” in *Proc. Int. Conf. Learn. Represent. (ICLR)*, New Orleans, LA, USA, May 2019.

[110] J. King and C. Meinhardt, “Privacy in an AI era: How do we protect our personal information?” Stanford HAI, Mar. 18, 2024. [Online]. Available: https://hai.stanford.edu/news/privacy-ai-era-how-do-we-protect-our-personal-information

[111] S. Karumbaiah, A. Lan, S. Nagpal, R. S. Baker, A. Botelho, and N. Heffernan, “Using past data to warm start active machine learning: Does context matter?,” in *Proc. 11th Int. Conf. Learn. Anal. Knowl. (LAK)*, Irvine, CA, USA, Apr. 2021, pp. 151–160.